# Coupled differential-algebraic equations framework for modeling six-degree-of-freedom flight dynamics of asymmetric fixed-wing aircraft

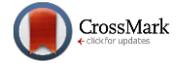

Osama A. Marzouk *

*College of Engineering, University of Buraimi, Al Buraimi, Oman*

## ABSTRACT

This study presents a comprehensive mathematical framework for modeling the flight dynamics of a six-degree-of-freedom fixed-wing aircraft as a rigid body with three control surfaces: rudder, elevators, and ailerons. The framework consists of 35 differential-algebraic equations (DAEs) and requires 30 constants to be specified. It supports both direct and inverse flight dynamics analyses. In direct dynamics, the historical profiles of control inputs (deflection angles and engine thrust) are specified, and the resulting flight trajectory is predicted. In inverse dynamics, the desired flight trajectory and an additional constraint are specified to determine the required control inputs. The framework employs wind axes for linear-momentum equations and body axes for angular-momentum equations, incorporates two flight path angles, and provides formulas for aerodynamic force and moment coefficients. Key advantages include improved computational efficiency, elimination of Euler angle singularities, and independence from symmetry assumptions with regard to the aircraft's moments of inertia. The model also accounts for nonlinear air density variations with altitude, up to 20 km above mean sea level, making it suitable for accurate and efficient flight dynamics simulations.

## 1. Introduction

Air transport is very important socially (for individuals) and commercially (for goods and globalized businesses) as a fast means to move items over long distances, especially across international borders (Dimitrios and Maria, 2018; Young, 2020; Zhang and Graham, 2020). According to statistics announced by the International Air Transport Association (IATA), there were about 4,200 million airline passengers in 2017 (before the COVID-19 restrictions on public transport), which is equivalent to more than one passenger in a commercial air trip per two inhabitants in the world; and the value of trade carried by air in 2017 exceeded 6 US$ trillion. The number of flights performed by the global airline industry in the same year was 36.4 million. IATA also announced that the airline revenues (both passenger and cargo services) in 2022 were about 720 US$ billion, which is close to the 2017 value, and 88% of the 2019 value (a pre-pandemic level). Air transport may expand from long-distance coverage to urban air mobility (UAM), allowing rapid commuting within a smart city (Kalakou et al., 2023; Marzouk, 2022). There are different types of aircraft, but the fixed-wing type dominates the global general aviation market, where this type has the advantage of a faster flight and higher efficiency (less consumption of energy per unit of flight time or flight distance) compared to the rotorcraft type (Crosby, 2022; Wang and Cai, 2015).

A prerequisite for the design of a fixed-wing aircraft, or for predicting the performance of another engineering system in general, is the ability to model its performance through simulations that involve solving a set of governing equations (Mi and Zhan, 2020; Rizzi, 2011; Marzouk and Nayfeh, 2010). Accurately modeling the flight dynamics of an aircraft plays a crucial role in advancing autonomous or remotely controlled unmanned aerial vehicles (UAVs) with reliable control. It also improves flight efficiency, reducing fuel consumption and emissions, leading to both economic and environmental benefits. Additionally, it enables the exploration of innovative designs and alternative propulsion systems for powering the aircraft during flight (Javaid et al., 2012; Tasca et al., 2021; Bravo-Mosquera et al., 2022; Marzouk and Nayfeh, 2009a; 2007).

The flight dynamics problem of an aircraft involves many interacting variables and effects, such as the aerodynamic loads, the aircraft attitude (orientation or tilt angles), the desired flight path (trajectory), and the onboard propulsion system (Durham, 2013). The flight dynamics problem may be divided into two categories. The first flight dynamics category is the direct (or forward or exploratory) problem, where the time history of the control variables (the deflection angles of movable control surfaces: ailerons, elevators, and rudder, and the engine thrust) are known, whereas the resulting travel path (the coordinates of the aircraft with respect to the initial take-off location) is unknown. This resembles the "direct kinematics" or "forward kinematics" problem in robotics when the joint position variables are known while the end-effector position and orientation are unknown (Cardona et al., 2024; Kang et al., 2012). The second flight dynamics category is the inverse (or normative) simulation problem, where a desired flight path (trajectory or maneuver) is known, while the time history of the control variables needed to accomplish this trajectory is unknown (Blajer et al., 2009; Gallup, 2023; Lu, 2010; Zhou and Wang, 2015). In either category, the equations of motion (EOMs) for the aircraft need to be solved to predict the time history of the unknown quantities.

Numerical simulation of a flight dynamics problem is a powerful design tool for aircraft that manifests the relationship between its input controls and output motion. Various design parameters (such as the wing area and the aerodynamic constants) may be tuned as a result of such simulation modeling, leading to an opportunity for computer-based optimization (Geske et al., 2024; Marzouk and Nayfeh, 2009b; 2008a; Laptev et al., 2024). For example, the extreme values of the required engine thrust force (thus, its deliverable propulsive power), control surface deflection angles, and the angle of attack for expected maneuvers that the aircraft may perform can be computed. These extreme values are then compared with the allowed values in the provisional design, and if the allowed limits are exceeded, then iterative modifications can be made. For example, wings have streamlined airfoil sections that permit a maximum airfoil's angle of attack (the angle of incidence between the airfoil section and the surrounding air) (Abbott, 2012). However, at some value of the airfoil's angle of attack, a stall phenomenon occurs where the flow around the airfoil is separated and becomes no longer streamlined enough, and large vortices develop that cause a decline in the lifting force. Through flight dynamics simulations, the designer can validate the suitability of the selected airfoil section for the expected maneuvers. If the stall is reached, then another airfoil section may be selected. As another example, if the target climb rate (increase in altitude with time) was found through flight dynamics simulations to require a higher level of power from the engine than the allowed limit, then the designer can modify the provisional design by sizing the engine for a higher power.

This article presents a detailed mathematical framework for modeling the flight dynamics of a fixed-wing airplane during general three-dimensional maneuvers, incorporating six degrees of freedom (6-DOFs). The nonlinear model comprises 35 differential-algebraic equations (DAEs) involving 39 variables and 30 auxiliary constant parameters. A differential-algebraic system, also known as an algebro-differential system, combines differential equations (which include derivatives) and algebraic equations (which relate variables without derivatives) (Lamour et al., 2013). Four variables need to be specified as inputs, while the remaining 35 variables can be obtained by solving the DAE system. In the case of direct (forward) simulation, the three control deflection angles (for the ailerons, elevators, and rudder) and the thrust force can be specified as either time-dependent functions or as numerical arrays of discrete values. In this case, the flight path coordinates with respect to the ground can be predicted. In the case of inverse simulation, the three trajectory coordinates are specified (as analytical functions or as discrete arrays), along with a fourth arbitrary flight variable; then, the four flight controls (three deflection angles and thrust) can be predicted.

While the topic of mathematical modeling of airplane's flight dynamics has been covered earlier elsewhere (Raol and Singh, 2023; Tai et al., 2023), the current study and the model presented herein have some advantages, as follows:

- The model handles six degrees of freedom (the most general motion for a rigid body).
- The model does not assume a plane of left-right symmetric (non-symmetric airplanes can be modeled).
- The model covers aerodynamic details (not just rigid body dynamics) with complete expressions for all the flight-dependent aerodynamic/stability coefficients.
- The model utilizes three sets of axes: ground axes (earth axes, inertial axes, or land axes), body-fixed axes (body axes), and wind axes. This is done instead of using the Euler angles directly to describe the attitude of the airplane (thus, its body-fixed axes) with respect to earth axes. The model utilizes two spherical (azimuth and elevation) flight-path angles as an intermediate system, allowing the separation of the flight-path angles from the airplane attitude angles (Euler angles), where both sets of angles can have very different values from each other. The model also utilizes two other spherical angles (the angle of attack and the sideslip angle) to describe the body axes relative to the velocity vector (and the wind axes), not with respect to the earth axes directly. It should be noted that the Euler angles (airplane attitude angles with respect to earth) are not eliminated. They are still needed in the linear-momentum equations. Also, their rates (first-time derivatives) are needed in the angular-momentum equations.

- The model simultaneously uses the wind axes (for the linear-momentum equations) and the body axes (for the angular-momentum equations), taking advantage of each set of axes rather than exclusively being limited to a single set throughout.
- The model captures the decline of air density with altitude (up to 20 km) rather than assuming it to be constant.
- The model provides the derivation or explanation for some equations, showing how they were obtained.
- The model is accompanied by several illustrating sketches to demonstrate axes systems or angles, which can be difficult to understand by formulas or textual descriptions alone (all these sketches were self-made, and no artificial intelligence tool was employed for generating them).
- The model is presented as detailed scalar equations rather than top-level vector equations or generic relations. Formulas are expanded to reveal the interrelation among variables and constant parameters. This makes the model useful particularly to readers interested in building a computational flight mechanics simulator.
- The model does not suffer from the singularity attributed to upward or downward flight (when flying exactly perpendicular to the horizon plan).

Despite these points of strength, the model has simplifying limitations, which help in bounding the complexity of the model. The model assumptions include:

- The gravitational acceleration is assumed to be constant (as the sea level value). At an altitude of $h$=20 km above sea level, the relative reduction in the gravitational acceleration is less than 1%. This is estimated as (Cavell et al., 2018).

% Reduction in gravitational acceleration at 20 km =
$$1 - \left(\frac{R_E}{R_E+h}\right)^2 = 1 - \left(\frac{6,371 \text{ km}}{6,391 \text{ km}}\right)^2 = 0.00625 \text{ (or } 0.625\%) \text{ (1.I)}$$

where, the mean earth radius is taken as $R_E$=6,371 km (Deng et al., 2008; Mahony, 2013).

- The aircraft mass is assumed to be constant during the flight. This assumption makes the model more accurate for smaller flights, which consume less fuel. This assumption is not applicable to battery-powered aircraft.
- The aircraft is treated as a single rigid body (with fixed dimensions and moments of inertia). Thus, the effect of the deflections on the control surfaces on these geometric parameters is ignored. The impact of simplification is not significant, given that the control surfaces are small relative to the airplane, and their deflections are structurally constrained.
- The thrust force is assumed to be concentrated along the longitudinal body axes. Distributed propulsion (Gohardani et al., 2011; Qiao et al., 2024) cannot be represented exactly. If the airplane has more than one engine (thus more than one source of propulsive thrust force), its resultant thrust vector should coincide with the longitudinal axis of the airplane and should be placed at the same distance from the airplane nose as the center of gravity of the airplane, such that they do not exert moments about the airplane's center of gravity (which is the origin of the airplane-fixed Cartesian axes).
- Other than the three types of control surfaces in the form of one aileron pair, one elevator pair, and one rudder, no additional movable surfaces are included. Extra movable elements such as wing flaps, leading edge slats, spoiler surfaces, trim tabs, and wing vortex generators are not addressed (Cole, 1990; Genç et al., 2009; Negahban et al., 2024; Pecora, 2021; Tian et al., 2017; Wang et al., 2019; Zajdel et al., 2022; 2023).
- The variation of the air density with altitude is represented up to an altitude of 20 km (65,616.8 ft). For benchmarking, typical cruise altitudes for conventional transport and commercial airplanes are between 30,000 to 42,000 feet (9.1 to 12.8 km), which is well within the covered range.

## 2. Structure of the model

In the coming sections, the 35 individual scalar equations that form together the 6-DOF DAEs (six-degree-of-freedom, differential-algebraic equations) model of flight dynamics for a general airplane are presented in the sequence given in Table 1.

**Table 1:** Differential-algebraic equations governing the 6-DOF motion of an airplane

| Group of equations | Number of equations | Differential or algebraic |
|---|---|---|
| Angular velocity vector in body axes | 3 | differential |
| Linear-momentum equations | 3 | differential |
| Angular-momentum equations | 6 | differential and algebraic |
| Transforming the linear velocity from spherical flight path axes to ground axes | 3 | differential |
| Relating flight path angles to aircraft attitude angles and wind angles | 2 | algebraic |
| Dynamic pressure and aerodynamic forces | 4 | algebraic |
| Moments | 3 | algebraic |
| Aerodynamic coefficients | 9 | algebraic |
| Air density as a function of altitude | 2 | algebraic |
| Total | 35 | - |

To differentiate the 35 equations in the DAE system from other equations, they are numbered sequentially using Arabic numerals (1 to 35). In contrast, other supplementary formulas, such as those used to explain specific quantities or derive expressions, are labeled with uppercase Roman numerals. These are preceded by a period and the corresponding section number where they first appear (e.g., 1.I, 2.I, 3.I, 3.II, 3.III, etc.). The 39 flight variables that appear in the DAE model are summarized in Table 2. Constant parameters that appear in the DAE system (such as the airplane moments of inertia about the center of gravity) are not considered flight variables because they do not change during the flight and are neither an input nor an output of the DAE system.

**Table 2:** Flight variables appear in the governing system of differential-algebraic equations

| Group of variables | Number of variables | Symbols |
|---|---|---|
| Euler angles and angular velocity components | 6 | $\phi, \theta, \psi, p, q, r$ |
| Spherical coordinates for the linear velocity | 3 | $V, \alpha, \beta$ |
| Flight path angles (spherical coordinates for the position) | 2 | $\theta_w, \psi_w$ |
| Airplane controls | 4 | $\delta_l, \delta_m, \delta_n, T$ |
| Inertial (ground-based) coordinates | 3 | $x_g, y_g, z_g$ |
| Dynamic pressure and aerodynamic forces | 4 | $\bar{q}, F_x, F_y, F_z$ |
| Auxiliary moments and total moments | 6 | $T_1, T_2, T_3, M_x, M_y, M_z$ |
| Aerodynamic coefficients | 9 | $C_L, C_D, C_C, C_x, C_y, C_z, C_l, C_m, C_n$ |
| Altitude and air density | 2 | $h, \rho$ |
| Total | 39 | - |

Also, the time derivative of a flight variable is not counted as a separate additional variable. Such differentiation processes are assumed to be readily possible either analytically (if the variable is known as a function of time and/or of other time-dependent variables) or numerically using the finite difference method (if the variable is known as a discrete time series of values) (Marzouk, 2009; 2010a; 2011b; Thomas, 2010; Zhang and Yao, 2013). A time derivative is designated by an overdot above the symbol. For example, $\dot{\theta}$ is the time derivative of $\theta$, or

$$\dot{\theta} = \frac{d\theta}{dt} \tag{2.I}$$

Components of a vector that can be easily computed using simple trigonometric rules are not considered separate variables. For example, the total velocity magnitude ($V$) is among the flight variables, but its body-axes components ($u, v, w$) are not among the flight variables because they are not totally independent quantities. In the presented DAE system, there is a singularity at zero aircraft velocity (hovering condition) because in the linear momentum equations (to be presented later), the time rate ($\dot{\beta}$) of the sideslip angle becomes undefined. There is another singularity at a sideslip angle value of $\pi/2$ or $-\pi/2$ (90° or -90°), because then $\cos\beta = 0$, and thus in the linear momentum equations (to be presented later), the time rate ($\dot{\alpha}$) of the angle of attack becomes undefined. Despite these two limitations, they are not of concern for fixed-wing airplanes, where a hovering condition and a condition of purely sideways travel are not expected. There is no singularity due to the second Euler angle taking the value $\pi/2$ or $-\pi/2$ (90° or -90°), because the terms $\tan\theta$ or $\sec\theta$ do not appear, which would cause the yaw rotation and the roll rotation to become indistinguishable, with the first Euler angle (or the yaw angle $\psi$) cannot be determined uniquely; and this is a feature known for the Euler angles known as the "gimbal lock" (Barman and Sinha, 2023a; 2023b; Brezov, 2024; De Paula et al., 2024; Han et al., 2023; Hanai et al., 2024; Hossain et al., 2024; Liu et al., 2023; Stankovic and Müller, 2024).

## 3. Angular velocity vector in body axes (three differential equations)

In the next section, it becomes clear that the angular velocity vector ($\vec{\Omega}_b$) referenced to the body axes is necessary for deriving the linear-momentum equations for the airplane. The purpose of the current section is to obtain three expressions for the three body-axes components ($p, q, r$) of the angular velocity vector in terms of aircraft attitude (Euler angles) and their time derivatives (Euler rates). We seek a representation for the angular velocity vector in the body axes with the following form:

$$\vec{\Omega}_b = p\,\hat{e}_{xb} + q\,\hat{e}_{yb} + r\,\hat{e}_{zb} \tag{3.I}$$

Let us consider three sequential rotations of the airplane (thus, rotations of its body-fixed axes) as shown in Fig. 1 when initially the airplane's body axes ($x_b, y_b, z_b$) are exactly aligned with the local earth/ground axes ($x_L, y_L z_L$). The first rotation is a yaw rotation about the local earth/ground axis ($z_L$), leading to an intermediate set of orthogonal axes ($x', y', z' = z_L$). The yaw rotation angle is the first Euler angle or the heading angle ($\psi$). This is followed by a second rotation that is a pitch rotation about the intermediate axis ($y'$), leading to another intimidate set of orthogonal axes ($x'', y'' = y', z''$). The pitch rotation angle is the second Euler angle ($\theta$). This is followed by a third rotation, that is a roll rotation about the intermediate axis ($x''$), leading eventually to the body axes ($x_b = x'', y_b, z_b$). The roll rotation angle is the third Euler angle or the bank angle ($\phi$).

The angular velocity vector can be expressed as three components in the three non-orthogonal axes about which the rotations took place, leading to

$$\vec{\Omega}_{b'L} = \dot{\phi}\,\hat{e}_{xb} + \dot{\theta}\,\hat{e}_{y'} + \dot{\psi}\,\hat{e}_{zL} \tag{3.II}$$

The intermediate unit vector ($\hat{e}_{y'}$) can be resolved into two orthogonal components in the body axes ($\hat{e}_{yb}, \hat{e}_{zb}$), as

$$\hat{e}_{y'} = \cos\phi\,\hat{e}_{yb} - \sin\phi\,\hat{e}_{zb}, \tag{3.III}$$

using Eq. 3.III into Eq. 3.II gives

$$\vec{\Omega}_{bL} = \dot{\phi}\,\hat{e}_{xb} + \cos\phi\,\dot{\theta}\,\hat{e}_{yb} - \sin\phi\,\dot{\theta}\,\hat{e}_{zb} + \dot{\psi}\,\hat{e}_{zL}. \qquad (3.IV)$$

The local earth unit vector ($\hat{e}_{zL}$) can be resolved into three orthogonal components in the body axes ($\hat{e}_{xb}, \hat{e}_{yb}, \hat{e}_{zb}$), as

$$\hat{e}_{zL} = -\sin\theta\,\hat{e}_{xb} + \cos\theta\sin\phi\,\hat{e}_{yb} + \cos\theta\cos\phi\,\hat{e}_{zb}, \quad (3.V)$$

using Eq. 3.V into Eq. 3.IV gives

$$\vec{\Omega}_b = \dot{\phi}\,\hat{e}_{xb} + \cos\phi\,\dot{\theta}\,\hat{e}_{yb} - \sin\phi\,\dot{\theta}\,\hat{e}_{xb} - \sin\phi\,\dot{\theta}\,\hat{e}_{zb} - \sin\theta\,\dot{\psi}\,\hat{e}_{xb} + \cos\theta\sin\phi\,\dot{\psi}\,\hat{e}_{yb} + \cos\theta\cos\phi\,\dot{\psi}\,\hat{e}_{zb}. \quad (3.VI)$$

Collecting terms with common unit vectors gives

$$\vec{\Omega}_b = (\dot{\phi} - \sin\theta\,\dot{\psi})\,\hat{e}_{xb} + (\cos\phi\,\dot{\theta} + \cos\theta\sin\phi\,\dot{\psi})\,\hat{e}_{yb} + (\cos\theta\cos\phi\,\dot{\psi} - \sin\phi\,\dot{\theta})\,\hat{e}_{zb}, \qquad (3.VII)$$

equating the components in Eq. 3.VII with their respective components in Eq. 3.I gives

$$p = \dot{\phi} - \sin\theta\,\dot{\psi} \qquad (1)$$
$$q = \cos\phi\,\dot{\theta} + \cos\theta\sin\phi\,\dot{\psi} \qquad (2)$$
$$r = \cos\theta\cos\phi\,\dot{\psi} - \sin\phi\,\dot{\theta} \qquad (3)$$

# 4. Three linear-momentum equations (three differential equations)

In formulating the linear-momentum equations for the airplane, we divide the forces acting on the airplane at its center of gravity) into three types: (1) the weight force ($m g_0$) acting in the direction of the third local earth axis ($z_L$) pointing from the airplane's center of gravity toward the earth's center, (2) the aerodynamic force components as projected along the three body axes ($X, Y, Z$), and (3) the engine thrust force ($T$) acting in the longitudinal body axis ($x_b$).

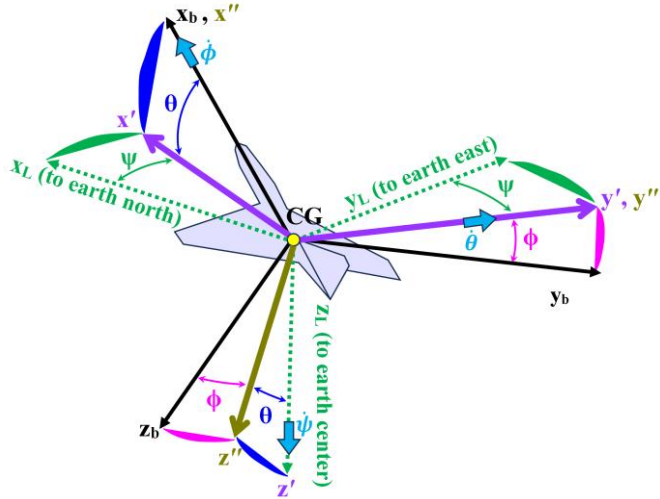

**Fig. 1:** Diagram of Euler rotations (yaw, pitch, roll) illustrating the transformation from local earth axes to body axes, with directions of positive angular rates ($\dot{\phi}, \dot{\theta}, \dot{\psi}$) indicated

If the translational equations of motion are formulated in the body axes, we obtain

$$m\left(\frac{d\vec{V}_b}{dt} + \vec{\Omega}_b \times \vec{V}_b\right) = \sum \vec{F}_b. \qquad (4.I)$$

The above vector equation can be expanded into the following three scalar equations:

$x_b$ component: $m(\dot{u} - v\,r + w\,q) = X - m g_0 \sin\theta + T$ (4.II)
$y_b$ component: $m(\dot{v} - w\,p + u\,r) = Y + m g_0 \cos\theta \sin\phi$ (4.III)
$z_b$ component: $m(\dot{w} - u\,q + v\,p) = Z + m g_0 \cos\theta \cos\phi$. (4.IV)

Fig. 2 illustrates the forces and moments acting on the airplane. Fig. 2 also shows the body-axes components ($u, v, w$) of the airplane's linear velocity (relative to the air, at the airplane's center of gravity) and the body-axes components ($p, q, r$) of the airplane's angular velocity. While, in principle, Eqs. 4.II and 4.IV can be retained as the equations of linear (translational) motion for the airplane, they may suffer from poor computational efficiency due to the possible large disparity in the magnitude of added/subtracted terms. As an example, we consider the case of a supersonic airplane (an airplane flying at a speed above the speed of sound) with a reasonable flight speed of 600 m/s (2,160 km/h), which is less than three times the speed of sound, or at a Mach number (the ratio between the flight speed and the local speed of sound) below 3 (Ao et al., 2023; Christian et al., 2023; Crocker, 1998; Della Posta et al., 2024; Fu et al., 2023b; Georgiadis et al., 2024; Huang et al., 2023; Huda and Edi, 2013; Li et al., 2023; Lui et al., 2024; Marzouk, 2008; 2020; Smith and Richards, 2023; Voet et al., 2024; Zamuraev and Kalinina, 2023). We also assume that this supersonic airplane has a reasonable upper limit on the pitch rate ($q$) of about 2 rad/s (or 114.6 degrees per second) (Costello and Jitprahai, 2002; Randall et al., 2012). In that presumed case, the artificial acceleration term ($u\,q$) in Eq. 4.IV can as be large as 1,200 m/s² or 122 g's. On the other hand, the force-induced acceleration term ($Z/m$) in the same equation may have an upper limit of only a few g's. Thus, the artificial accelerations can be greater than the actual accelerations by two orders of magnitude due to the high rotation rates experienced by the airplane (thus, by the body axes). This results in an unfavorable discrepancy of scales and may harmfully impact the solution accuracy for a given computer precision. In addition, Eqs. 4.II and 4.IV strongly couple the high-speed dynamics of rotation into the translational motion, which places severe computational demands.

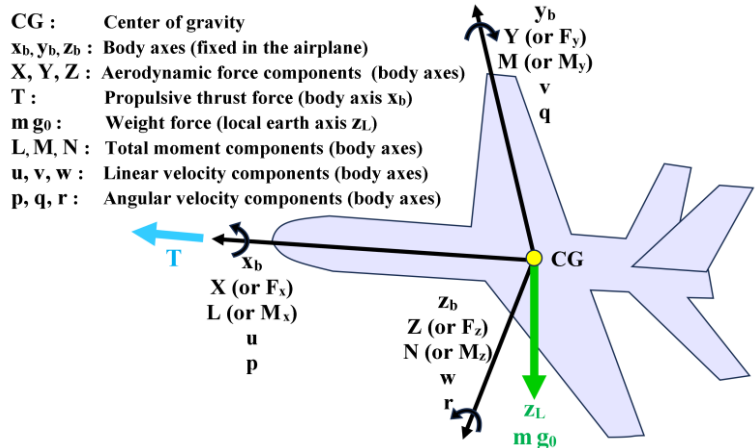

CG : Center of gravity
$x_b, y_b, z_b$ : Body axes (fixed in the airplane)
X, Y, Z : Aerodynamic force components (body axes)
T : Propulsive thrust force (body axis $x_b$)
m $g_0$ : Weight force (local earth axis $z_L$)
L, M, N : Total moment components (body axes)
u, v, w : Linear velocity components (body axes)
p, q, r : Angular velocity components (body axes)

**Fig. 2:** Illustration of the body axes and the components of various vectors along them

To address the above-mentioned issues, the translational equations are transformed from the body axes ($x_b, y_b, z_b$) to the wind axes ($x_w, y_w, z_w$), which are aligned with the flight-path tangent, which is the direction of the total velocity vector of the airplane relative to the air. For performing such transformation, the following geometric equalities are used (Figs. 3-6 provide a visual explanation of some angles that appear in the transformed equations).

$$V = \sqrt{u^2 + v^2 + w^2} \tag{4.V}$$
$$\beta = \sin^{-1}(v/V) \tag{4.VI}$$
$$\alpha = \tan^{-1}(w/u) \tag{4.VII}$$

Thus, we have

$$\sin \beta = v/V \tag{4.VIII}$$
$$\cos \beta = \sqrt{u^2 + w^2}/V \tag{4.IX}$$
$$\tan \beta = v/\sqrt{u^2 + w^2} \tag{4.X}$$
$$\tan \alpha = w/u \tag{4.XI}$$
$$\sin \alpha = w/\sqrt{u^2 + w^2} \tag{4.XII}$$
$$\cos \alpha = u/\sqrt{u^2 + w^2} \tag{4.XIII}$$
$$\cos \alpha \cos \beta = u/V \tag{4.XIV}$$
$$\sin \alpha \cos \beta = w/V. \tag{4.XV}$$

Therefore, each body-axis velocity component can be related to the total velocity magnitude ($V$) through the spherical angles ($\alpha$) and/or ($\beta$) as follows:

$$u = V \cos \alpha \cos \beta \tag{4.XVI}$$
$$v = V \sin \beta \tag{4.XVII}$$
$$w = V \sin \alpha \cos \beta. \tag{4.XVIII}$$

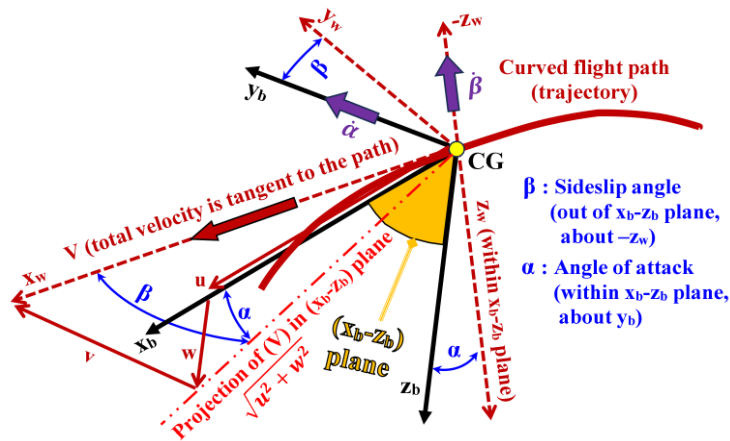

**Fig. 3:** Diagram illustrating the spherical angles ($\alpha, \beta$) for transformation from body axes to wind axes, including directions of positive angular rates ($\dot{\alpha}, \dot{\beta}$) and body-axes velocity components ($u, v, w$)

This angle of attack ($\alpha$) lies in the body plane $x_b - z_b$, between the body axis ($x_b$) and the projected total velocity vector in that plane. It is positive when the airplane nose is above the flight path tangent (the projected total velocity vector lies in the quadrant between the positive $x_b$, and the positive $z_b$), causing an upward lifting force. In the current work, the angle of attack is defined as a single angle at the whole airplane level (not defined at the smaller level of the wing or wing's airfoil sections).

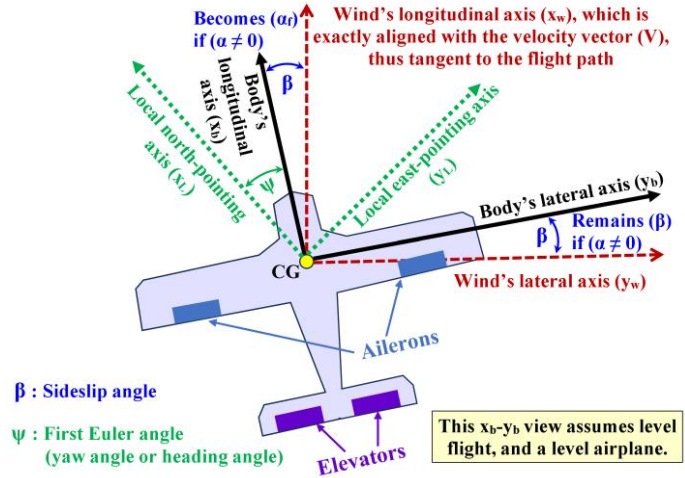

**Fig. 4:** Illustration of the flight dynamics in a planar view (body-fixed plane $x_b - y_b$)

The flank angle of attack ($\alpha_f$) lies in the body plane $x_b - y_b$, making it easy to visualize by looking at that body-fixed plane. When the aircraft is viewed from above, the body plan $x_b - y_b$, the angle between the line of the total velocity ($V$) and its lateral component ($v$) is the flank angle of attack ($\alpha_f$), which is a projection of the sideslip angle ($\beta$). In this case, if there is no angle of attack ($\alpha = 0$), then ($\beta$) becomes identical to ($\alpha_f$). The flank angle of attack ($\alpha_f$) is defined as not one of the flight variables in the presented DAE system. It is not needed for the flight dynamics framework of this

article. However, its value during the flight can be obtained and reported as an extra output according to

$$\alpha_f = \tan^{-1}(v/u). \tag{4.XIX}$$

In addition, the following relations apply:

$$\tan \beta = \tan \alpha_f \cos \alpha = v/\sqrt{u^2 + w^2} \tag{4.XX}$$
$$\tan \alpha_f = \tan \beta / \cos \alpha = v/u \tag{4.XXI}$$
$$\beta = \tan^{-1}(\tan \alpha_f \cos \alpha) = \tan^{-1}(v/\sqrt{u^2 + w^2}) \tag{4.XXII}$$
$$\alpha_f = \tan^{-1}(\tan \beta / \cos \alpha) = \tan^{-1}(v/u). \tag{4.XXIII}$$

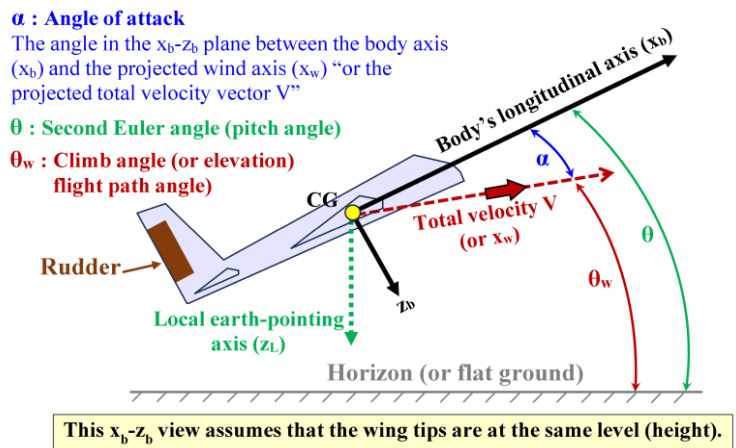

**Fig. 5:** Illustration of the flight dynamics in a planar view (body-fixed plane $x_b - z_b$)

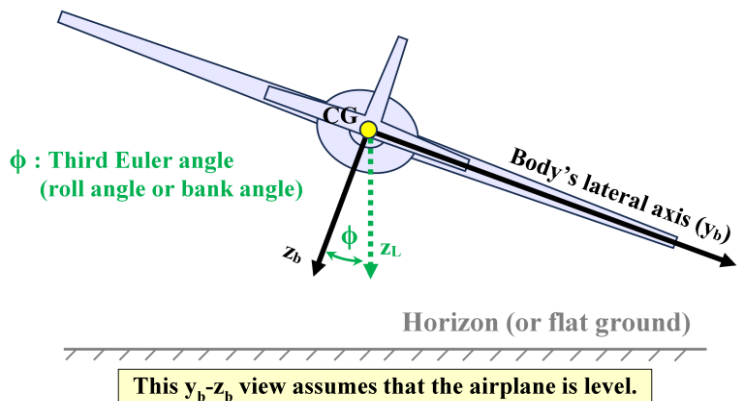

**Fig. 6:** Illustration of the flight dynamics in a planar view (body-fixed plane $y_b - z_b$)

The final transformed equations of translational motion (equations of linear momentum) are:

$x_w$ component:
$$m\,\dot{V} = \bar{q}\,S\left(C_x \cos\alpha\cos\beta + C_y \sin\beta + C_z \sin\alpha\cos\beta\right) + m\,g_0(\cos\theta\sin\phi\sin\alpha\cos\beta - \sin\theta\cos\alpha\cos\beta + \cos\theta\cos\phi\sin\alpha\cos\beta) + T\cos\alpha\cos\beta \tag{4}$$

$y_w$ component:
$$m\,V\,\dot{\beta} = \bar{q}\,S\left(C_y \cos\beta - C_x \cos\alpha\sin\beta - C_z \sin\alpha\sin\beta\right) + m\,g_0(\cos\theta\sin\phi\sin\beta + \sin\theta\cos\alpha\sin\beta - \cos\theta\cos\phi\sin\alpha\sin\beta) + T\cos\alpha\sin\beta + m\,V\,(-r\cos\alpha + p\sin\alpha) \tag{5}$$

$z_w$ component:
$$m\,V\cos\beta\,\dot{\alpha} = \bar{q}\,S\left(C_z \cos\alpha - C_x \sin\alpha\right) + m\,g_0(\sin\theta\sin\alpha + \cos\theta\cos\phi\cos\alpha) - T\sin\alpha + m\,V\,(q\cos\beta - r\sin\alpha\sin\beta - p\cos\alpha\sin\beta) \tag{6}$$

It can be noticed that the first Euler angle, the heading angle ($\psi$) does not appear in the above three equations of translational motion. This is explained by the fact that the direction in which the airplane's nose is pointing relative to the ground during flight does not impact the dynamics of its motion. The heading angle ($\psi$) relates the airplane's orientation with respect to an earth axis, which is arbitrarily chosen.

## 5. Angular-momentum equations (three differential equations, three algebraic equations)

Unlike the translational equations of motion (EOMs), which were formulated in the wind axes, the rotational equations of motion (or the equations of angular momentum) are to be formulated in the body axes. This is highly advantageous because the moments of inertia can then be treated as constant geometric parameters rather than time-varying quantities.

Before presenting the differential equations of angular momentum, a derived geometric constant needs to be obtained. It is the determinant of the inertia matrix (tensor) (Melnikov, 2012; Rucker and Wensing, 2022). It is designated here by the symbol ($T_0$). It is related to the individual moments of inertia (about the body axes) as follows:

$$T_0 = \begin{vmatrix} A & -F & -E \\ -F & B & -D \\ -E & -D & C \end{vmatrix} = \begin{vmatrix} I_{xx} & -I_{xy} & -I_{xz} \\ -I_{xy} & I_{yy} & -I_{yz} \\ -I_{xz} & -I_{yz} & I_{zz} \end{vmatrix} = A\,B\,C - A\,D^2 - B\,E^2 - C\,F^2 - 2\,D\,E\,F \tag{5.I}$$

where, the operator | | means the determinant.

It should be noted that while the above equation for the inertia constant ($T_0$) is important. It is not counted as one of the DAE equations because it involves geometric inertia constants only. Thus, the constant ($T_0$) is not a flight variable but a constant parameter that is computed only once from the known six elements of the inertia tensor.

Then, three auxiliary moments are defined through three algebraic equations (which are parts of the DAE system). These auxiliary moments are used in formulating the three main differential equations for the rotational motion of the airplane and enable writing them efficiently in a relatively compact form. The equations for obtaining auxiliary moments are

$$T_1 = (B - C)\,q\,r + (E\,q - F\,r)\,p + (q^2 - r^2)\,D + L \tag{7}$$
$$T_2 = (C - A)\,r\,p + (F\,r - D\,p)\,q + (r^2 - p^2)\,E + M \tag{8}$$
$$T_3 = (A - B)\,p\,q + (D\,p - E\,q)\,r + (p^2 - q^2)\,F + N \tag{9}$$

Finally, the main translational equations of motion (as differential equations that are also parts of the DAE system) about the body axes are

$x_b$ component :
$$T_0\,\dot{p} = (B\,C - D^2)\,T_1 + (F\,C + E\,D)\,T_2 + (F\,D + E\,B)\,T_3 \tag{10}$$
$y_b$ component :
$$T_0\,\dot{q} = (A\,C - E^2)\,T_2 + (A\,D + E\,F)\,T_3 + (F\,C + E\,D)\,T_1 \tag{11}$$
$z_b$ component :
$$T_0\,\dot{r} = (A\,B - F^2)\,T_3 + (F\,D + B\,E)\,T_1 + (A\,D + F\,E)\,T_2. \tag{12}$$

It is worth mentioning that in the special case of a symmetric airplane (with the body-fixed plane $x_b - z_b$ being a plane of symmetry), the mass products of inertia ($D = I_{yz}, F = I_{xy}$) vanish (Lorenzetti et al., 2017). For this special case, the overall translational equations of motion can be reduced from six equations to three equations. Our presented DAE model is generic (not assuming any geometric symmetry). However, the three reduced differential angular-momentum equations are provided below as additional information, which can be used instead of the previous six equations (three algebraic equations and three differential equations) for airplanes with a plane of symmetry (the left/port half is a reflected version of the right/starboard half) (Beknalkar et al., 2024; Braca et al., 2014; Kim et al., 2024; Rizzi et al., 2024; Svozil, 2023; Tewari, 2016).

Reduced angular momentum equation, $x_b$ component: $\dot{p}\,(A\,C - E^2) = (B\,C - E^2 - C^2)\,q\,r + (A - B + C)\,E\,p\,q + C\,L + E\,N$ (5.II)
or : $\dot{p}\,(I_{xx}\,I_{zz} - I_{xz}^2) = (I_{yy}\,I_{zz} - I_{zz}^2 - I_{zz}^2)\,q\,r + (I_{xx} - I_{yy} + I_{zz})\,I_{xz}\,p\,q + I_{zz}\,M_x + I_{xz}\,M_z$ (5.III)
Reduced angular momentum equation, $y_b$ component: $\dot{q}\,B = E\,r^2 - E\,p^2 + (C - A)\,p\,r + M$ (5.IV)
or : $\dot{q}\,I_{yy} = I_{xz}\,r^2 - I_{xz}\,p^2 + (I_{zz} - I_{xx})\,p\,r + M_y$ (5.V)
Reduced angular momentum equation, $z_b$ component: $\dot{r}\,(A^2 + E^2 - A\,B)\,p\,q + (B - A - C)\,E\,q\,r + A\,N + E\,L$ (5.VI)
or : $\dot{r}\,(I_{xx}\,I_{zz} - I_{xz}^2) = (I_{xx}^2 + I_{xz}^2 - I_{xx}\,I_{yy})\,p\,q + (I_{yy} - I_{xx} - I_{xx})\,I_{xz}\,q\,r + I_{xx}\,M_z + I_{xz}\,M_x$ (5.VII)

## 6. Transforming the linear velocity from spherical flight path axes to ground axes (three algebraic equations)

Fig. 7 explains the meaning of the spherical angles ($\theta_w, \psi_w$), where ($\psi_w$) is the azimuth angle (as of a compass reading) of the projected line on the ground for the straight line connecting the airplane to its take-off point, and ($\theta_w$) is the elevation angle of that straight line above the horizon. The climb angle or the elevation flight path angle ($\theta_w$) may also be called "first flight path angle." The lateral flight path angle or the azimuth flight path angle ($\psi_w$) may also be called "second flight path angle." The radial coordinate of this spherical coordinate system is the straight-line distance between the moving airplane and the fixed take-off point (which is the origin of this spherical coordinate system).

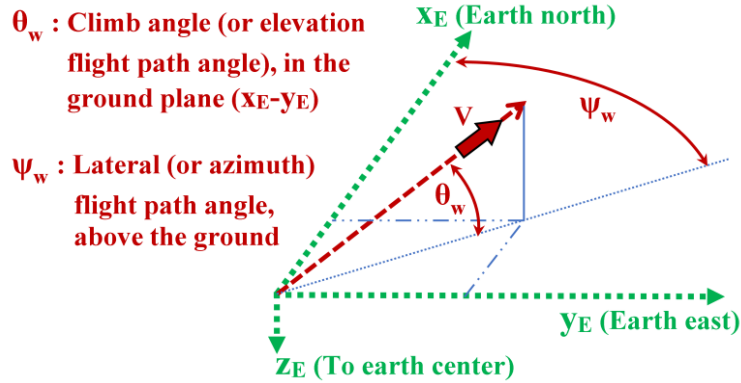

**Fig. 7:** Illustration of the two flight path angles

Fig. 8 demonstrates the meaning of the flight path coordinates ($x_g, y_g, z_g$). Fig. 8 also shows two of the three airplane attitude angles (Euler angles), namely the yaw angle ($\psi$) and the pitch angle ($\theta$). Comparing these two Euler angles to the flight path angles ($\theta_w, \psi_w$) in the Fig. 7 helps clarify the difference between the two pairs of angles. It should be noted that the flight path angles are independent of the airplane orientation (attitude), while the Euler angles are independent of the flight path (location of the airplane in space).

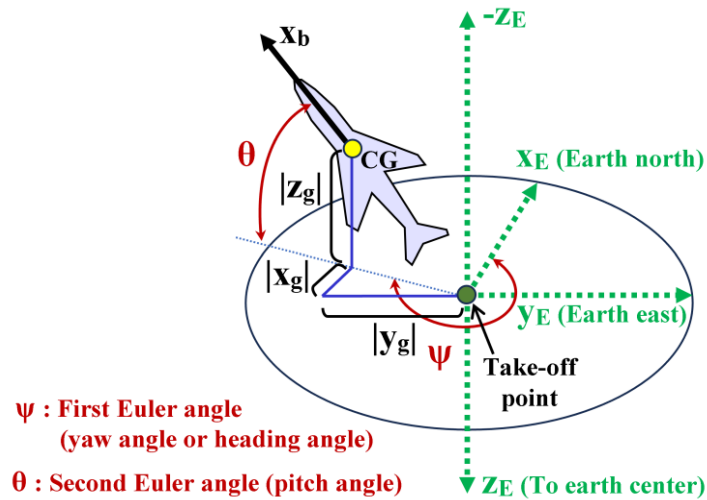

**Fig. 8:** Visualization of flight path coordinates from the take-off point, showing all three coordinates with negative values in the depicted position. The operator | | denotes absolute value (magnitude)

The equations relating the rates of the inertial coordinates (ground-referenced Cartesian coordinates, measured from the take-off point at the start of the flight) of the airplane trajectory ($\dot{x}_g, \dot{y}_g, \dot{z}_g$) to the total velocity magnitude and the flight path angles derived from the straightforward resolution of the velocity vector (with a magnitude $V$) into its components along the earth axes (either the aircraft-local earth axes $x_L, y_L, z_L$, or their parallel ground-fixed inertial earth axes $x_E, y_E, z_E$). These equations are

$$\dot{x}_g = V \cos\theta_w \cos\psi_w \tag{13}$$
$$\dot{y}_g = V \cos\theta_w \sin\psi_w \tag{14}$$
$$\dot{z}_g = -V \sin\theta_w \tag{15}$$

It is worth mentioning that the climb rate ($\dot{h}$) (which is the time rate of the altitude) is the negative of the time rate of the third flight path coordinate of the flight trajectory, or ($z_g$). This is because ($z_g$) increases when the airplane descends toward the ground, while the altitude ($h$) increases when the airplane ascends away from the ground. Thus,

$$\dot{h} = -\dot{z}_g \tag{6.I}$$

From Eqs. 15 and 6.I, the climb rate ($\dot{h}$) can be obtained as

$$\dot{h} = V \sin\theta_w \tag{6.II}$$

## 7. Relating flight path angles to aircraft attitude angles and wind angles (two algebraic equations)

The DAE system of equations requires two additional algebraic equations that relate the two flight path angles to other angular flight variables, particularly the three Euler angles ($\boldsymbol{\phi}, \boldsymbol{\theta}, \boldsymbol{\psi}$), the angle of attack ($\boldsymbol{\alpha}$), and the sideslip angle ($\boldsymbol{\beta}$). These equations are

$$\cos\theta_w \sin(\psi_w - \psi) = \cos\phi \sin\beta - \sin\phi \sin\alpha \cos\beta \quad (16)$$
$$\sin\theta_w = \sin\theta \cos\alpha \cos\beta - \cos\theta \sin\phi \sin\beta - \cos\theta \cos\phi \sin\alpha \cos\beta \quad (17)$$

The last equation, Eq. 17, can be derived by noticing that ($\sin\theta_w$) is the result of dividing the climb rate ($\dot{h}$), as an opposite side in a right-angled velocity triangle, by the total velocity magnitude ($V$), as the hypotenuse in that triangle. Alternatively, Eq. 6.II can show this as follows:

$$\dot{h}/V = V \sin\theta_w/V = \sin\theta_w \quad (7.I)$$

The climb rate ($\dot{h}$) can be related to the body-axes velocity components ($u, v, w$) as

$$\dot{h} = u \sin\theta - v \cos\theta \sin\phi - w \cos\theta \cos\phi \quad (7.II)$$

Recalling Eqs. 4.XVI, 4.XVII, and 4.XVIII, these body-axes velocity components can, in turn, be expressed in terms of the total velocity magnitude ($V$) as ($u = V \cos\alpha \cos\beta$), ($v = V \sin\beta$), and ($w = V \sin\alpha \cos\beta$); respectively. Using these expressions in Eq. 7.II gives

$$\dot{h} = (V \cos\alpha \cos\beta)(\sin\theta) - (V \sin\beta)(\cos\theta \sin\phi) - (V \sin\alpha \cos\beta)(\cos\theta \cos\phi). \quad (7.III)$$

Dividing both sides by $V$, and recalling that $\dot{h}/V = \sin\theta_w$ gives

$$\sin\theta_w = \cos\alpha \cos\beta \sin\theta - \sin\beta \cos\theta \sin\phi - \sin\alpha \cos\beta \cos\theta \cos\phi, \quad (7.IV)$$

which is exactly equivalent to Eq. 17.

It may also be useful to add that in the special case where the total velocity vector is aligned with the longitudinal body axis ($x_b$), thus the body axes coincide with the wind axes ($\alpha = 0, \beta = 0$), then the two airplane attitude angles or Euler angles ($\psi, \theta$) become identical to the flight path angles ($\psi_w, \theta_w$). This can also be inferred from Eqs. 16 and 17. However, the remaining third airplane attitude angle for the roll ($\phi$), or the third Euler angle, can still take arbitrary values in this special case.

## 8. Dynamic pressure and aerodynamic forces (four algebraic equations)

The aerodynamic force vector acting on the simulated airplane is most easily handled in the body-fixed axes. Here, we refer to the three body-axes components of the aerodynamic effects (the interaction between the air and the airplane moving through it) as three aerodynamic forces, but the reader should recognize that they are not three independent forces but projected components.

These aerodynamic forces are functions of nondimensional aerodynamic coefficients that should be obtained for the airplane. As typically done in the aeronautical field, the nondimensional aerodynamic force coefficients ($C_x, C_y, C_z$) are converted into forces using the dynamic pressure ($\bar{q}$), which depends on the air density and the magnitude of the flight velocity (relative to air) and a reference area ($S$) that is the projected (planform) area of the wing, which is a constant geometric parameter for the airplane (Qin et al., 2016; Marzouk and Nayfeh, 2008b; Secco and Mattos, 2017; Song et al., 2012). The following expressions describe the dynamic pressure and the body-axes aerodynamic forces that act on the airplane:

$$\bar{q} = 0.5 \, \rho \, V^2 \quad (18)$$
$$x_b \text{ component: } X \text{ (or } F_x) = \bar{q} \, S \, C_x \quad (19)$$
$$y_b \text{ component: } Y \text{ (or } F_y) = \bar{q} \, S \, C_y \quad (20)$$
$$z_b \text{ component: } Z \text{ (or } F_z) = \bar{q} \, S \, C_z \quad (21)$$

As noted earlier (in Section 4, Three Linear-Momentum Equations), the aerodynamic forces are not the only forces acting on the airplane. Two additional forces include the weight and the thrust (the forward propulsive force generated by the airplane's engine). This section focuses exclusively on the aerodynamic forces, as the weight and thrust forces are fundamental and do not require mathematical definitions.

## 9. Moments (three algebraic equations)

The total moment vector is resolved into three components along the body axes, which we refer to as three moments. These moments are expressed in terms of nondimensional moment coefficients ($C_l, C_m, C_n$), the dynamic pressure, the reference surface area of the wing, and the reference length. For the pitching moment, the mean wing chord (the average straight-line distance from the leading edge to the trailing edge of the wing) can be used as the reference length. Such a reference length is assigned here as the symbol ($c$) (Meku et al., 2023; Spedding and McArthur, 2010). For the rolling and yawing moments, the wingspan (straight distance from the left/port tip to the right/starboard tip of the wing) may be used as a reference length. Such a reference length is assigned here the symbol $b$ (Custodio et al., 2015; Santos et al., 2017).

The following expressions describe the moments that act on the airplane:

$$x_b \text{ component: } L \text{ (or } M_x) = \bar{q} \, S \, b \, C_l \quad (22)$$
$$y_b \text{ component: } M \text{ (or } M_y) = \bar{q} \, S \, c \, C_m \quad (23)$$
$$z_b \text{ component: } N \text{ (or } M_z) = \bar{q} \, S \, b \, C_n. \quad (24)$$

We clarify here that the moments of the body axes in the DAE system arise only from the aerodynamic effects. Because the thrust acts along the body axis ($x_b$) and thus, its direction vector passes through the origin of the body axes, and the airplane weight is at the origin of the body axes. These two non-aerodynamic forces do not cause additional moments.

## 10. Aerodynamic coefficients (nine algebraic equations)

There are nine nondimensional aerodynamic/stability coefficients involved in the DAE system, related to three force components parallel to the wind axes, three force components in the positive body axes, and three-moment components along the body axes.

In the section 8, it was shown that three nondimensional aerodynamic-force coefficients need to be found as a part of the DAE system for modeling the flight dynamics of the airplane. These coefficients are $C_x, C_y, C_z$, which allows determining the three corresponding components of the aerodynamic force vector in the body axes. These three force components are $X$ or $F_x, Y$ or $F_y, Z$ or $F_z$. Ideally, one should attempt to find the aerodynamic coefficients ($C_x, C_y, C_z$) directly from other flight variables. However, in aeronautics, there is another set of nondimensional aerodynamic-force coefficients that are commonly reported for airplanes, and these coefficients can be estimated using specialized aeronautical software or through testing of a small-scale prototype in a controlled environment. Thus, this other set is obtained first, and from them, the more relevant coefficients ($C_x, C_y, C_z$) are deduced. The preparatory set of aerodynamic coefficients is the lift coefficient ($C_L$) and the drag coefficient ($C_D$), which can be viewed as normalized (nondimensionalized) versions of the lift and drag, respectively (Aslanov, 2017; Marzouk, 2011a; 2010b; Islas-Narvaez et al., 2022; Jung and Rezgui, 2023; Zheng et al., 2024). In addition, there is a third preparatory aerodynamic coefficient, which is the side-force coefficient ($C_C$). These preparatory aerodynamic coefficients are related to the components of the aerodynamic force vector when resolved in the wind axes, rather than in the body axes. The drag component of the aerodynamic force vector (or simply the drag) is in the opposite direction of the total velocity vector (thus, it opposes the airplane's translation motion). The lift component of the aerodynamic force vector (or simply the lift) is perpendicular to the drag force, and the upward part of the lift that is perpendicular to the horizon is responsible for counteracting the gravitational effect. The side component of the aerodynamic force vector (or simply the side force) is perpendicular to both the drag and the lift.

The non-dimensional lift coefficient ($C_L$) is approximated here as a linear function of the angle of attack ($\alpha$), with a constant slope ($C_{L\alpha} = dC_L/d\alpha$) that should be known for the airplane (Qu et al., 2015). It should be noted that while this simple equation is often used by aeronautical engineers, it breaks down remarkably at a critical, relatively high angle of attack (typically near 15° or 0.26 rad) due to the separation of the boundary layer from the wing skin, accompanied with the formation of undesired vortices at the upper side of the wing, leading to a sharp decline in the lift coefficient after reaching a maximum value of ($C_{Lmax}$); and this phenomenon is known as a wing stall condition (Andreu Angulo and Ansell, 2019; Geissler and van der Wall, 2017; Hammer et al., 2022). If the nonlinear region of the lift coefficient function may be encountered, then the linear model for the $C_L(\alpha)$ function should simply be replaced with an extended nonlinear one, with no other modifications needed in the DAE system. For many airfoil sections of wings, the lift coefficient has a non-zero value (denoted by $C_{L0}$) at a zero angle of attack. This intercept value is due to the asymmetry (slight curvature or camber) of the airfoil section, while for symmetric airfoils, $C_{L0} = 0$ (Xu and Lagor, 2021). We consider the general case of cambered airfoil (and $C_{L0}$ should be known for the airplane as an input constant parameter), and thus the relation between the lift coefficient and the angle of attack is modeled as

$$C_L = C_{L0} + C_{L\alpha}\, \alpha. \tag{25}$$

Eq. 25 is graphically demonstrated in Fig. 9.

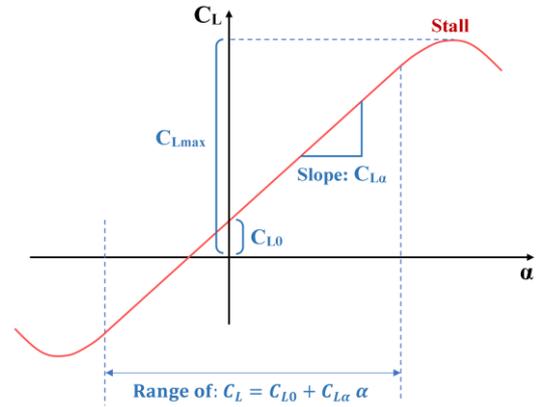

**Fig. 9:** Illustration of the relation between the lift coefficient and angle of attack, sowing the stall region, and the region of linear dependence

The drag coefficient ($C_D$) is modeled as a quadratic function of the lift coefficient through a law called the drag-polar relationship (Hantrais-Gervois and Destarac, 2015; Sun et al., 2020). The drag coefficient equation requires that two aerodynamic constants ($C_{D0}, K_{CD}$) are provided as input aerodynamic parameters for the airplane. The drag coefficient equation is

$$C_D = C_{D0} + K_{CD}\, C_L^2 \tag{26}$$

A side-force coefficient ($C_C$) is modeled as a linear function of the sideslip angle ($\beta$), with the constant of proportionality designed by ($C_{C\beta}$). Thus, we have

$$C_C = C_{C\beta}\, \beta \tag{27}$$

The remaining aerodynamic force coefficients (with respect to body axes) are expressed as

$$C_x = -C_D \cos\alpha \cos\beta - C_C \cos\alpha \sin\beta + C_L \sin\alpha \tag{28}$$
$$C_y = -C_D \sin\beta + C_C \cos\beta \tag{29}$$

$$C_z = -C_D \sin\alpha \cos\beta - C_C \sin\alpha \sin\beta - C_L \cos\alpha \qquad (30)$$

The total-moment coefficients (with respect to body axes) are modeled such that they do not only depend on flight angles but also on body-fixed angular rates and on the deflection angles of the three control surfaces. There are 14 constant aerodynamic and stability constants introduced in the three expressions for the three-moment coefficients, which are $(C_{l\beta}, C_{lp}, C_{lr}, C_{l\delta l}, C_{l\delta n})$ for the dimensionless rolling-moment coefficient $(C_l)$; $(C_{m0}, C_{m\alpha}, C_{mq}, C_{m\delta m})$, for the dimensionless pitching-moment coefficient $(C_m)$; and $(C_{n\beta}, C_{np}, C_{nr}, C_{n\delta l}, C_{n\delta n})$ for the dimensionless yawing-moment coefficient $(C_n)$. These constants should be known for the airplane as input parameters. It can be noticed that there is a strong coupling between the rolling and yawing. This is a known feature in fixed-wing aircraft (Brincklow et al., 2024; Stengel, 2022). For example, a rudder deflection intended to cause a yaw rotation also causes a rolling moment, while an aileron's deflection intended to cause a roll rotation also causes a yawing moment.

$$C_l = C_{l\beta}\beta + C_{lp} p\,(b/V) + C_{lr} r\,(b/V) + C_{l\delta l}\,\delta_l + C_{l\delta n}\,\delta_n \qquad (31)$$

$$C_m = C_{m0} + C_{m\alpha}\alpha + C_{mq} q\,(c/V) + C_{m\delta m}\,\delta_m \qquad (32)$$

$$C_n = C_{n\beta}\beta + C_{np} p\,(b/V) + C_{nr} r\,(b/V) + C_{n\delta l}\,\delta_l + C_{n\delta n}\,\delta_n \qquad (33)$$

As a remark, the multiplication of the pitch rate $q$ by $(c/V)$, the multiplication of the roll rate $(p)$ by $(b/V)$, and the multiplication of the yaw rate $(r)$ by $(b/V)$ makes these scaled angular rates nondimensional, and thus, they can be properly multiplied by the respective constant parameters (called stability derivatives) in the above three equations.

As an additional remark, Eqs. 31-33 are formulated in a form that suits a forward (direct) simulation mode, where the control deflections are known and thus appear on the right-hand side. They can be manipulated to suit the inverse simulation mode, where the deflection angles are unknown and need to be found. For example, the manipulated version of Eq. 32 for inverse simulation becomes:

$$\delta_m = (C_m - C_{m0} - C_{m\alpha}\,\alpha - C_{mq}\,q\,c/V)/C_{m\delta m} \qquad (10.I)$$

The angle $(\delta_m)$ is the deflection angle of either of the two movable elevators that are used for pitching and for longitudinal stability. This angle is positive if the trailing edges of the elevators move concurrently down (thus causing the aircraft to make a positive pitch angle, with the nose tilting up).

Due to the coupling between Eqs. 31 and 33, they need to be manipulated simultaneously as a system of two algebraic equations in two unknowns $(\delta_l, \delta_n)$. Solving these two deflection angles gives

$$\delta_l = \left(C_{n\delta n}\left(C_l - C_{l\beta}\beta - C_{lp}\,pb/V - C_{lr}\,rb/V\right) - C_{l\delta n}\left(C_n - C_{n\beta}\beta - C_{np}\,pb/V - C_{nr}\,rb/V\right)\right)/(C_{l\delta l}\,C_{n\delta n} - C_{l\delta n}\,C_{n\delta l}) \qquad (10.II)$$

$$\delta_n = \left(C_{l\delta l}\left(C_n - C_{n\beta}\beta - C_{np}\,pb/V - C_{nr}\,rb/V\right) - C_{n\delta l}\left(C_l - C_{l\beta}\beta - C_{lp}\,pb/V - C_{lr}\,rb/V\right)\right)/(C_{l\delta l}\,C_{n\delta n} - C_{l\delta n}\,C_{n\delta l}) \qquad (10.III)$$

The angle $\delta_l$ is the deflection angle of either of the two movable ailerons that are used for rolling and lateral control. This angle is positive if the trailing edge of the right/starboard aileron moves up while the trailing edge of the left/port aileron moves down (thus causing the aircraft to make a positive roll angle, with the left/port side of the wing tilting up).

The angle $\delta_n$ is the deflection angle of the movable rudder that is used for yawing and directional control. This angle is positive if the trailing edge of the rudder moves toward the left/port side of the aircraft and the pilot (thus causing the aircraft to make a positive yaw angle, with the nose tilting left).

## 11. Air Density as a function of altitude (two algebraic equations)

The last part of the DAE system for the flight dynamics framework is the computation of the air density as a function of altitude (above the global mean sea level). Unlike gravitational acceleration, which decreases only very weakly with altitude (thus, it is treated here as constant at the sea level value), the air density declines much faster as the airplane climbs to high altitudes due to the decline of the air pressure because of the smaller air column (whose weight induces this air pressure) in the remaining outer part of the atmosphere.

The variation of the air density with the altitude is governed here by the International Standard Atmosphere (ISA) model (Kuprikov, 2023; Santana et al., 2019). In the ISA model, the air is assumed to be a dry mixture (free from water vapor), with a mole fraction of 78.09% for nitrogen, 20.95% for oxygen, and 0.93% for argon, and traces of other constituent gases such as carbon dioxide, neon, helium, krypton, and hydrogen (Marzouk, 2023).

It should be noted that the ISA model uses a gravity-adjusted version of the altitude called the "geopotential altitude," whereas we use the ordinary "true" geometric altitude that is a directly measured height above the mean sea level, MSL (Atasay et al., 2023; Ferreira and Gimeno, 2024; Kitajima et al., 2024; Levchenko and Levchenko, 2023; Matsui et al., 2023a; 2023b; Matyja et al., 2023; Oksuztepe et al., 2023; Stober et al., 2020; Tamesue et al., 2023; Yuan et al., 2024). This gravity adjustment aims to account for the impact of decreasing gravitational acceleration with altitude, but this effect is neglected in our flight dynamics model. The "virtual" geopotential altitude is an adjusted value that is slightly less than the "true" geometric altitude to accommodate for the effect of a slowly decreasing gravitational acceleration, thus reducing the weight of the outer air column. To obtain the geopotential altitude ($H$) that corresponds to a particular geometric altitude ($h$), the following equation can be used (Scherllin-Pirscher et al., 2017):

$$H = \frac{1}{g_0} \int_{\zeta=0}^{\zeta=h} \tilde{g}(\zeta)\, d\zeta \qquad (11.I)$$

where, $\zeta$ is a dummy integration variable that represents the geometric altitude, and the function ($\tilde{g}$) describes the decline in the gravitational acceleration with the geometric altitude ($h$) as

$$\tilde{g}(h) = g_0 \left(\frac{R_E}{R_E + h}\right)^2 \qquad (11.II)$$

where, $R_E$ is the mean earth's radius, taken as 6,371,000 m (Orlov, 2018).

At a geometric altitude of 20,000 m (which is the upper limit on altitudes in the presented framework), the corresponding geopotential altitude is 19,937 m, which means that the difference is 63 m, which represents a relative discrepancy of only 0.315% (relative to the geometric value of 20,000 m) or 0.316% (relative to the geopotential value of 19,937 m).

The geometric altitude ($h$) above the sea level can be obtained from the third flight path coordinate ($z_g$) and the initial geometric altitude ($h_{ini}$) at the start of the flight, at the take-off point as

$$h = h_{ini} - z_g \qquad (34)$$

It should be noted that ($z_g$) takes negative values as the airplane ascends. Thus, $h > h_{ini}$ if the airplane is at a higher position relative to the take-off point.

The part of the atmosphere covered here for air density variation is the troposphere layer (from 0 m to 11,000 m above sea level) and the subsequent tropopause layer (from 11,000 m to 20,000 m above sea level), which are together sufficient for accommodating a wide range of aviation activities. In the troposphere, the air temperature is assumed (according to ISA) to decline regularly with altitude at a fixed rate (lapse rate) of $\lambda$=0.0065 K/m, making it a non-isothermal layer. The air is treated as an ideal compressible gas with a specific gas constant of $R$=287.05 J/kg.K (Marzouk, 2017; Woody, 2013). The standard sea level air density is $\rho_0$=1.225 kg/m³, and the standard sea level temperature is $\Theta_0$=288.15 K (equivalent to 15.00 °C) (Gerkema and Duran-Matute, 2017). The gravitational acceleration (treated as a constant) is $g_0$=9.80665 m/s². From the principles of thermodynamics and fluid mechanics, the nonlinear decline in the air density with the altitude above sea level in the non-isothermal troposphere layer can be derived (Bar-Meir, 2022; Struchtrup, 2014).

$$\rho = \rho_0 \left(1 - (\lambda/\Theta_0)\, h\right)^{g_0/(R\,\lambda)-1} \qquad (11.III)$$

For the above equation, we assign the symbol ($m_0$) to the constant ($\lambda/\Theta_0$) that multiplies the altitude ($h$) with $m_0$=2.25577 × 10⁻⁵ m⁻¹. We also assign the symbol ($n_0$) to the constant nondimensional exponent ($g_0/(R\,\lambda)-1$), with $n_0$=4.25593.

Thus, the air density as a function of altitude in the troposphere (the closest layer to the ground) can be rewritten as

$$\rho = \rho_0 \left(1 - m_0 h\right)^{n_0} = 1.225 \left(1 - 2.25577 \times 10^{-5}\, h\right)^{4.25593} \qquad (11.IV)$$

Where the output air density is kg/m3 and the altitude is m. This troposphere expression is also applicable at negative altitudes (which may occur at deep ground depression located beneath the mean seal level).

The higher tropopause layer has a constant temperature of $\Theta_1$=216.65 K (equivalent to −56.50 °C), and this layer starts (according to the ISA model) at an altitude of $h_1$=11,000 m. The air density at this transition altitude, as computed from Eq. 11.IV, is $\rho_1$=0.36391 kg/m³. From the principles of thermodynamics and fluid mechanics, the nonlinear decline in the air density with the altitude above the sea level (mean sea level) in the isothermal tropopause layer can be derived as

$$\rho = \rho_1 \exp\!\left(-(g_0/R\,\Theta_1)\,(h - h_1)\right) \qquad (11.V)$$

where, **exp** is the exponential function (Marzouk, 2018).

In the above equation, we assign the symbol ($m_1$) to the constant ($g_0/(R\,\Theta_1)$) that multiplies the excess altitude ($h - h_1$) beyond the beginning of the tropopause layer, with $m_1$=1.57690×10⁻⁴ m⁻¹.

Thus, the air density as a function of altitude in the tropopause (the second closest layer to the ground) can be rewritten as

$$\rho = \rho_1 \exp(-m_1 (h - h_1)) = 0.36391 \exp(-1.57690 \times 10^{-4} (h - 11{,}000)) \qquad (11.VI)$$

Combining Eqs. 11.IV and 11.VI, A single formula can be written for the air density as a function of altitude for either layer of atmosphere we covered, and this is the last equation among the differential-algebraic equation systems for flight dynamics (DAE system). This equation is

$$\rho(h) = \begin{cases} \rho_0 \left(1 - m_0 h\right)^{n_0} & h \leq 11{,}000 \text{ m} \\ \rho_1 \exp\!\left(-m_1 (h - h_1)\right) & 11{,}000 \text{ m} < h \leq 20{,}000 \text{ m} \end{cases} \qquad (35)$$

Table 3 lists computed values of the air density according to Eq. 35, for altitudes up to 20,000 m; with an increment of 1,000 m. Table 3 also lists the air density ratio ($\sigma$), which is the ratio of the air density at a given altitude to the standard sea level density or

$$\sigma = \rho(h)/\rho_0 = \rho(h)/(1.225 \text{ kg/m}^3) \qquad (11.VII)$$

Table 3 also lists the air temperatures at these selected altitudes according to the International Standard Atmosphere (ISA) model. The air temperature is of interest in aviation because the speed of sound ($a$) in air is proportional to the square root of this temperature, and the speed of sound is an important speed threshold for objects moving in air, with $a$=340.29 m/s (or 1,225 km/h) at sea level (Chen et al., 2006; Hay, 2013; Jameel et al., 2021; Torenbeek, 2013). Specifically, the relationship between the speed of sound and the absolute temperature (the temperature expressed in kelvins) is (Guo et al., 2023; Hu et al., 2016; Hwang et al., 2019).

$$a = \sqrt{\gamma R \theta} \tag{11.VIII}$$

where, $\gamma$ is the ratio of specific heats (also called the adiabatic index, the isentropic exponent, or the specific heat ratio), which is approximately 1.40 for air near room temperature (Abbaszadehmosayebi and Ganippa, 2014; Fix et al., 2024; Palit et al., 2019; Prashantha et al., 2023; Simpson et al., 2023). From Table 3, it is seen that the density drops remarkably at high altitudes, reaching about 30% of its sea-level value and the end of the troposphere, and drops further to about 7% of the sea-level value at the end of the tropopause.

**Table 3:** Atmospheric conditions at different altitudes, according to the International Standard Atmosphere (ISA)

| Atmospheric layer | Altitude, $h$ (m) | Density, $\rho$ (kg/m³) | Density ratio, $\sigma$ (-) | Temperature, $\theta$ (K) |
|---|---|---|---|---|
| | 0 | 1.22500 | 1.00000 | 288.15 (15.00 °C) |
| | 1,000 | 1.11164 | 0.90746 | 281.65 (8.50 °C) |
| | 2,000 | 1.00649 | 0.82162 | 275.15 (2.00 °C) |
| | 3,000 | 0.90912 | 0.74214 | 268.65 (−4.50 °C) |
| | 4,000 | 0.81913 | 0.66867 | 262.15 (−11.00 °C) |
| Troposphere | 5,000 | 0.73611 | 0.60091 | 255.65 (−17.50 °C) |
| | 6,000 | 0.65969 | 0.53852 | 249.15 (−24.00 °C) |
| | 7,000 | 0.58950 | 0.48122 | 242.65 (−30.50 °C) |
| | 8,000 | 0.52516 | 0.42870 | 236.15 (−37.00 °C) |
| | 9,000 | 0.46634 | 0.38069 | 229.65 (−43.50 °C) |
| | 10,000 | 0.41270 | 0.33690 | 223.15 (−50.00 °C) |
| | 11,000 | 0.36391 | 0.29707 | 216.65 (−56.50 °C) |
| | 12,000 | 0.31082 | 0.25373 | 216.65 (−56.50 °C) |
| | 13,000 | 0.26548 | 0.21672 | 216.65 (−56.50 °C) |
| | 14,000 | 0.22675 | 0.18510 | 216.65 (−56.50 °C) |
| | 15,000 | 0.19367 | 0.15810 | 216.65 (−56.50 °C) |
| Tropopause | 16,000 | 0.16542 | 0.13503 | 216.65 (−56.50 °C) |
| | 17,0000 | 0.14128 | 0.11533 | 216.65 (−56.50 °C) |
| | 18,000 | 0.12067 | 0.09851 | 216.65 (−56.50 °C) |
| | 19,000 | 0.10307 | 0.08414 | 216.65 (−56.50 °C) |
| | 20,000 | 0.08803 | 0.07186 | 216.65 (−56.50 °C) |

It may be useful to add here that the absolute pressure, $P$ (the true, unadjusted pressure measured from the absolute vacuum reference) of air at a given altitude can be estimated from the absolute temperature and the computed air density at the same altitude, using one of the following forms of the ideal gas law

$$P(h) = \rho(h) \, R \, \theta(h) \tag{11.IX}$$

where, the $P$ is in pascals (Pa), $\rho$ is in kg/m³, and the air's gas constant $R$ is in J/kg.K (Akasaka et al., 2023; Das and Chatterjee, 2023; Duben, 2024; Fu et al., 2023a; Garrett, 2020; Kaushik, 2022; Khaji et al., 2023; Nandagopal, 2023; 2024).

## 12. Results and discussion

As the current study aimed at presenting a comprehensive mathematical framework for modeling the flight dynamics of a general fixed-wing airplane, the results are in the form of a set of 35 differential-algebraic equations (DAE system) that govern various aspects (precisely 39 flight variables) during a flight maneuver. These equations were divided into groups and were presented in the preceding nine sections (from section 3 to section 11). In the current section, we discuss two aspects of the model. First, the actual implementation of the model in a computerized flight simulator requires the provision of 30 constants (29 design parameters for the airplane and one trajectory parameter). These needed constant parameters are

- Airplane mass ($m$)
- Wing planform area ($S$)
- Reference longitudinal length, such as the mean chord ($c$)
- Reference lateral length, such as the wing span ($b$)
- Six mass moments of inertia about body axes ($A, B, C, D, E, F$)
- Five aerodynamic-force constants ($C_{L0}, C_{L\alpha}, C_{D0}, K_{CD}, C_{C\beta}$)
- Fourteen stability derivatives ($C_{l\beta}, C_{lp}, C_{lr}, C_{l\delta l}, C_{l\delta n}, C_{m0}, C_{m\alpha}, C_{mq}, C_{m\delta m}, C_{n\beta}, , C_{nr}, C_{n\delta l}, C_{n\delta n}$)
- Initial altitude (above sea level), at the take-off point ($h_{ini}$)

Second, in the simplest flight condition, which is steady level flight without acceleration or rotation (thus, the flight variables $\alpha, \beta, \theta, \phi, p, q, r$ vanish, as well as the time derivatives $\dot{V}, \dot{\alpha}, \dot{\beta}$), the presented DAE system automatically reduces to only two non-trivial equations, which derive from the first and third equations of linear momentum along the wind axes (which, in this special case, coincide with the body axes and even with the local earth axes). Eq. 4, which is the first differential linear-momentum equation (the first equation of translational motion), reduces to a simple algebraic force balance as

$$T = \bar{q}\,S\,C_D \qquad (12.I)$$

which states that the horizontal propulsive thrust exactly offsets the horizontal resistive drag. Eq. 6, which is the third differential linear-momentum equation (the third equation of translational motion), also reduces to simple algebraic force equality as

$$m\,g_0 = \bar{q}\,S\,C_L \qquad (12.II)$$

which states that the upward lift force exactly offsets the downward weight.

## 13. Conclusions

We presented a detailed framework for modeling the flight dynamics of a generic fixed-wing airplane, allowing for six-degree-of-freedom motion. The overall framework is a coupled nonlinear system consisting of 35 scalar equations (a mix of 12 differential equations and 23 algebraic equations), and it was listed in an organized and logical way, supported by some derivation and visual illustrations. The variation of the air density with altitude is captured up to a reasonable altitude of 20 km, following a set of assumptions that corresponds to the ISA.

The model includes 39 flight variables and 30 user-defined constant parameters. Therefore, four variables need to be specified as functional constraints to enable obtaining a unique solution. The selection of these constraints depends on whether the framework is to be solved as a forward (direct) simulation tool or as an inverse (normative) simulation tool.

In the case of forward (direct) simulation, the temporal profiles of four controls (thrust, ailerons' deflection angle, elevators' deflection angle, and rudder deflection angle) should be specified, and then the model explores what travel path results from these input profiles.

In the case of inverse (normative) simulation, the temporal profiles of the three coordinates of the target flight trajectory and a fourth arbitrary flight variable (such as pre-determined evolution of the bank/roll angle or the sideslip angle) should be specified, and then the model estimates how the four control variables need to change simultaneously to satisfy the desired trajectory.

## List of symbols

| | |
|---|---|
| $A$ (or $I_{xx}$) | Mass moments of inertia about the body-axis $x_b$ [kg.m2] |
| $a$ | Speed of sound in air [m/s] |
| $B$ (or $I_{yy}$) | Mass moments of inertia about the body-axis $y_b$ [kg.m2] |
| $b$ | Reference length for lateral stability (roll) or directional stability (yaw); can be the wing span [m] |
| (or $I_{zz}$) $C$ | Mass moments of inertia about the body-axis $z_b$ [kg.m2] |
| $c$ | Reference length for longitudinal stability; can be the wing mean chord [m] |
| $C_C$ | Side-force coefficient. The aerodynamic side force acts in the side wind axis $y_w$ [dimensionless] |
| $C_{C\beta}$ | Aerodynamic constant that describes the linear dependence of the side-force coefficient on the sideslip angle [1/rad] |
| $C_D$ | Drag coefficient. The drag acts in the negative direction of the flight-path-tangent wind axis $x_w$ [dimensionless] |
| $C_{D0}$ | Zero-lift drag coefficient (or parasitic drag coefficient). It is the constant part of the drag coefficient, not dependent on the exerted lift [dimensionless] |
| $C_L$ | Lift coefficient. The lift acts in the negative direction of the third wind axis $z_w$ [dimensionless] |
| $C_{L0}$ | Lift coefficient at zero angle of attack [dimensionless] |
| $C_{L\alpha}$ | Derivative of the lift coefficient with respect to the angle of attack [1/rad] |
| $C_l$ | Rolling-moment coefficient [dimensionless] |
| $C_{lp}$ | Aerodynamic constant that describes the linear dependence of the rolling-moment coefficient on the nondimensionalized roll rate [dimensionless] |
| $C_{lr}$ | Aerodynamic constant that describes the linear dependence of the rolling-moment coefficient on the nondimensionalized yaw rate [dimensionless] |
| $C_{l\beta}$ | Aerodynamic constant that describes the linear dependence of the rolling-moment coefficient on the sideslip angle [1/rad] |
| $C_{l\delta l}$ | Aerodynamic constant that describes the linear dependence of the rolling-moment coefficient on the ailerons' deflection angle [1/rad] |
| $C_{l\delta n}$ | Aerodynamic constant that describes the linear dependence of the rolling-moment coefficient on the rudder deflection angle [1/rad] |
| $C_m$ | Pitching-moment coefficient [dimensionless] |
| $C_{m0}$ | Aerodynamic constant, which is a fixed (parasitic) part of the pitching-moment coefficient, if applicable [dimensionless] |
| $C_{mq}$ | Aerodynamic constant that describes the linear dependence of the pitching-moment coefficient on the nondimensionalized pitch rate [dimensionless] |
| $C_{m\alpha}$ | Aerodynamic constant that describes the linear dependence of the pitching-moment coefficient on the angle of attack [1/rad] |
| $C_{m\delta m}$ | Aerodynamic constant that describes the linear dependence of the pitching-moment coefficient on the elevators' deflection angle [1/rad] |
| $C_n$ | Yawing-moment coefficient [dimensionless] |
| $C_{np}$ | Aerodynamic constant that describes the linear dependence of the yawing-moment coefficient on the nondimensionalized roll rate [dimensionless] |
| $C_{nr}$ | Aerodynamic constant that describes the linear dependence of the yawing-moment coefficient on the nondimensionalized yaw rate [dimensionless] |
| $C_{n\beta}$ | Aerodynamic constant that describes the linear dependence of the yawing-moment coefficient on the sideslip angle [1/rad] |
| $C_{n\delta l}$ | Aerodynamic constant that describes the linear dependence of the yawing-moment coefficient on the ailerons' deflection angle [1/rad] |
| $C_{n\delta n}$ | Aerodynamic constant that describes the linear dependence of the yawing-moment coefficient on the rudder deflection angle [1/rad] |
| $C_x$ | Aerodynamic force coefficient in the body axis $x_b$ [dimensionless] |
| $C_y$ | Aerodynamic force coefficient in the body axis $y_b$ [dimensionless] |
| $C_z$ | Aerodynamic force coefficient in the body axis $z_b$ [dimensionless] |
| CG | Center of gravity (center of mass, or centroid) of the aircraft |
| $D$ (or $I_{yz}$) | Mass product of inertia in the body-axes plane $y_b - z_b$ [kg.m2] |
| $E$ (or $I_{xz}$) | Mass product of inertia in the body-axes plane $x_b - z_b$ [kg.m2] |
| $\hat{e}_{xb}$ | Unit vector in the body axis $x_b$ |
| $\hat{e}_{yb}$ | Unit vector in the body axis $y_b$ |
| $\hat{e}_{zb}$ | Unit vector in the body axis $z_b$ |
| $\hat{e}_{y'}$ | Unit vector in the temporary rotation axis $y'$ |
| $\hat{e}_{zL}$ | Unit vector in the earth (ground) axis $z_L$ |
| $F$ (or $I_{xy}$) | Mass product of inertia in the body-axes plane $x_b - y_b$ [kg.m2] |
| $\sum \vec{F}_b$ | Force vector (accounting for all forces, aerodynamic, weight, and thrust), with components expressed in the body axes |
| or $g_0 g$ | Gravitational acceleration; treated as a constant at the sea level value [9.80665 m/s2] |
| $H$ | Geopotential altitude above the sea level [m] |
| $h$ | Geometric altitude above the sea level. The altitudes in the current study are geometric (true) altitudes by default, rather than geopotential altitudes. [m] |
| $h_1$ | Altitude above the sea level, at the transition between the troposphere layer and the tropopause layer, according to the International Standard Atmosphere (ISA) model [11,000 m] |
| $h_{ini}$ | Geometric altitude above the sea level at the take-off point [m] |
| $K_{CD}$ | Lift-induced drag coefficient. It is a constant that describes the quadratic dependence of the drag coefficient on the lift coefficient. [dimensionless] |
| $L$ (or $M_x$) | Rolling moment about the body axis $x_b$. It is positive if the left (port) side of the wing rises up relative to the right (starboard) side [N.m] |
| $M$ (or $M_y$) | Pitching moment about the body axis $y_b$. It is positive if the aircraft nose (front tip) rises up relative to the tail [N.m] |
| $m$ | Mass of the aircraft [kg] |
| $m_0$ | Derived constant for use in the nonlinear equation relating the air density to the altitude in the troposphere layer, according to the International Standard Atmosphere (ISA) model, $\lambda/\theta_0$ [2.25577E-5 1/m] |
| $m_1$ | Derived constant for use in the nonlinear equation relating the air density to the altitude in the tropopause layer, according to the International Standard Atmosphere (ISA) model, $g_0/(R\,\Theta_1)$ [1.57690E-4 1/m] |
| $N$ (or $M_z$) | Yawing moment about the body axis $z_b$. It is |
| | positive if the aircraft nose (front tip) rotates toward the right (starboard) side [N.m] |
| $n_0$ | Derived constant exponent for use in the nonlinear equation relating the air density to the altitude in the troposphere layer, according to the International Standard Atmosphere (ISA) model, $g_0/(R\,\lambda) - 1$ [4.25593, dimensionless] |
| $P$ | Absolute pressure of air [Pa or N/m2] |
| $p$ | Angular velocity of the rolling rotation (or roll rate), about the body axis $x_b$ [rad/s] |
| $q$ | Angular velocity of the pitching rotation (or pitch rate), about the body axis $y_b$ [rad/s] |
| $\bar{q}$ | Dynamic pressure (or velocity pressure), $\bar{q} = 0.5\,\rho\,V^2$ [N/m2 or Pa "pascal"] |
| $R$ | Gas constant of the air, when treated as a homogenous ideal gas [287.05 J/kg.K] |
| $R_E$ | Earth's mean radius [6,371,000 m] |
| $r$ | Angular velocity of the yawing rotation (or yaw rate), about the body axis $z_b$ [rad/s] |
| $S$ | Wing planform (projected) area [m2] |
| $T$ | Magnitude of the thrust, which is treated here as purely directed toward the body axis $x_b$ [N "newton"] |
| $T_0$ | Geometric constant derived from the aircraft's moments of inertia. It is the determinant of the inertia matrix (tensor of inertia) [kg3.m6] |
| $T_1$ | Auxiliary moment quantity used in the angular-momentum equations [kg.m2/s2 or N.m] |
| $T_2$ | Auxiliary moment quantity used in the angular-momentum equations [kg.m2/s2 or N.m] |
| $T_3$ | Auxiliary moment quantity used in the angular-momentum equations [kg.m2/s2 or N.m] |
| $t$ | Time [s "second"] |
| $u$ | Component of the total velocity vector (at the aircraft's center of gravity) in the body axis $x_b$ [m/s] |
| $V$ | Total velocity (magnitude) of the aircraft's center of gravity relative to the surrounding air [m/s] |
| $\vec{V}_b$ | Total linear velocity vector (of the center of gravity), with components expressed in the body axes |
| $v$ | Component of the total velocity vector (at the aircraft's center of gravity) in the body axis $y_b$ [m/s] |
| $w$ | Component of the total velocity vector (at the aircraft's center of gravity) in the body axis $z_b$ [m/s] |
| $X$ (or $F_x$) | Component of the aerodynamic force vector (weight and thrust excluded) in the body axis $x_b$ [N] |
| $x_b$ | Longitudinal body axis of the aircraft (fixed in the aircraft, from its center of gravity to its nose) |
| $x_E$ | Inertial or global earth (ground) axis aligned with the geographic north, and fixed to the ground at the take-off point |
| $x_g$ | Component of the flight path (trajectory) in the inertial earth (ground) axis $x_E$. This is the net (positive minus negative) distance traveled in the geographic north, measured from the initial take-off point [m] |
| $x_L$ | Local earth (ground) axis that is always aligned with the geographic north (as an invariant reference direction), but it is translated with the airplane (fixed at its center of gravity) |
| $x_w$ | Flight wind axis, which is always tangent to the flight path (trajectory). It is in the same |

direction as the total velocity vector. Its origin is the aircraft's center of gravity. The drag force is always in the negative direction of $x_w$

$Y$ (or $F_y$) | Component of the aerodynamic force vector (weight and thrust excluded) in the body axis $y_b$ [N]

$y'$ | Temporary body-fixed axis, that is the right/starboard body-axis $y_b$ after a yaw rotation about the body-axis $z_b$

$y_b$ | Lateral body axis of the aircraft (fixed in the aircraft, from its center of gravity to its right/starboard side)

$y_E$ | Inertial or global earth (ground) axis aligned with the geographic east, and fixed to the ground at the take-off point

$y_g$ | Component of the flight path (trajectory) in the inertial earth (ground) axis $y_E$. This is the net (positive minus negative) distance traveled in the geographic east, measured from the initial take-off point [m]

$y_L$ | Local earth (ground) axis that is always aligned with the geographic east (as an invariant reference direction), but it is translated with the airplane (fixed at its center of gravity)

$y_w$ | Side wind axis, which is perpendicular to the flight path. Its direction is uniquely determined such that it is aligned with the local east earth axis $y_L$ when the other wind axis $x_w$ is aligned with the local-north earth axis $x_L$. Its origin is the aircraft's center of gravity.

$Z$ (or $F_z$) | Component of the aerodynamic force vector (weight and thrust excluded) in the body axis $z_b$ [N]

$z_b$ | Downward (or floor) body axis of the aircraft (fixed in the aircraft, from its center of gravity to its bottom/floor)

$z_E$ | Inertial or global earth (ground) axis pointing toward the earth's center, and fixed to the ground at the take-off point

$z_g$ | Component of the flight path (trajectory) in the inertial earth (ground) axis $z_E$. This is the net (positive minus negative) distance traveled toward the earth's center, measured from the initial take-off point [m]

$z_L$ | Local earth (ground) axis that is always pointing to the earth's center (as an invariant reference direction), but it is translated with the airplane (fixed at its center of gravity). It is always aligned with the gravitational acceleration

$z_w$ | Third wind axis, which is perpendicular to both the flight path tangent (which coincides with the wind axis $x_w$) and the wind axis $y_w$. Its origin is the aircraft's center of gravity. The lift force is always in the negative direction of $z_w$

$\alpha$ | Angle of attack (AOA); $\tan^{-1}(w/u)$ [rad]

$\alpha_f$ | Flank angle of attack; $\tan^{-1}(v/u)$ [rad]

$\beta$ | Sideslip angle or angle of side-slip (AOSS); $\sin^{-1}(v/V)$ [rad]

$\gamma$ | Specific heat ratio (or adiabatic index) for air [dimensionless]

$\delta_l$ | Deflection angle of either of the two movable ailerons that are used for rolling [rad]

$\delta_m$ | Deflection angle of either of the two movable elevators that are used for pitching [rad]

$\delta_n$ | Deflection angle of the movable rudder that is used for yawing [rad]

$\phi$ | Roll angle or bank angle; one of the three aircraft attitude angles. It represents a rolling rotation about the body axis $x_b$ (if no other rotations are made simultaneously). It is positive if the left/port side of the aircraft tilts down relative to the right/starboard side. It is also called the third Euler angle when interpreted in the context of rotational translations from inertial axes to the body-fixed axes. The range is: $-\pi \leq \phi \leq \pi$ [rad]

$\rho$ | Air density, depends on the altitude [kg/m3]

$\lambda$ | Lapse rate in the troposphere layer (decrease in temperature per unit increase in altitude). This is a constant according to the International Standard Atmosphere (ISA) model [0.0065 K/m]

$\rho_0$ | Air density at the sea level, according to the International Standard Atmosphere (ISA) model [1.2250 kg/m3]

$\rho_1$ | Air density at the transition between the troposphere layer and the tropopause layer, located at an altitude of 11,000 m according to the International Standard Atmosphere (ISA) model [0.36391 kg/m3]

$\sigma$ | Air density ratio (relative to the standard value at the sea level) [dimensionless]

$\Theta_0$ | Air temperature at the sea level, according to the International Standard Atmosphere (ISA) model [288.15 K "equivalent to 15.00 °C"]

$\Theta_1$ | Air temperature at the transition between the troposphere layer and the tropopause layer and within the entire isothermal (constant temperature) tropopause layer, according to the International Standard Atmosphere (ISA) model [216.65 K "equivalent to –56.50 °C"]

$\theta$ | Pitch angle; one of the three aircraft attitude angles. It represents a pitching rotation about the body axis $y_b$ (if no other rotations are made simultaneously). It is positive if the nose of the aircraft tilts up. It is also called the second Euler angle when interpreted in the context of rotational translations from inertial axes to the body-fixed axes. The range is: $-\pi/2 \leq \theta \leq \pi/2$. [rad]

$\theta_w$ | Climb angle, elevation flight path angle, or first flight path angle; one of the two flight path angles used to describe the flight path (trajectory) in spherical axes. It represents the angle between the horizon (or flat ground) and the straight line connecting the aircraft to the take-off point. The range is: $0 \leq \theta_w \leq \pi/2$. [rad]. If the flight path is level (no change in altitude) then $\theta_w = 0$

$\vec{\Omega}_b$ | Angular velocity vector for the aircraft as a rigid body, with the components expressed entirely in the body axes

$\vec{\Omega}_{bL}$ | Angular velocity vector for the aircraft as a rigid body, with the components expressed partly in the three body axes and partly in the local earth axis ($z_L$)

$\vec{\Omega}_{b'L}$ | Angular velocity vector for the aircraft as a rigid body, with the components expressed in three non-orthogonal axes representing three sequential rotations (roll about $z_L$, then pitch about $y'$, then yaw about $x_b$)

$\psi$ | Yaw angle or heading angle; one of the three aircraft attitude angles. It represents a yawing rotation about the body axis $z_b$ (if no other rotations are made simultaneously). It is positive if the nose of the aircraft tilts toward the right/starboard side. It also is called the

first Euler angle when interpreted in the context of rotational translations from inertial axes to the body-fixed axes. The range is: $-\pi \leq \psi \leq \pi$ [rad]

$\psi_w$    Lateral (or azimuth) flight path angle, or second flight path angle; one of the two flight path angles used to describe the flight path (trajectory) in spherical axes. It represents the angle in the horizon (or in the earth plane $x_E - y_E$) between the ground axis $x_E$ (the earth's north) and projection of the straight line connecting the aircraft to the take-off point. The range is: $0 \leq \psi_w \leq 2\pi$. If the flight path is due east, then $\psi_w = \pi/2$. If the flight path is due south, then $\psi_w = \pi$. If the flight path is due west, then $\psi_w = 3\pi/2$. [rad]

$\zeta$    Dummy integration variable that represents the geometric altitude above the sea level [m]

## Compliance with ethical standards

## Conflict of interest

The author declares no conflict of interest in preparing and executing this study. The research was conducted with complete adherence to academic integrity and objectivity.